\def\rfr#1{eq. (\ref{#1})}
\def\rfrs#1#2{(\ref{#1})-(\ref{#2})}

\def\dert#1#2{\frac{{{d}}{#1}}{{{d}}{#2}}}              

\def\rdm{\rho_{\rm dm}}

\def\bar{\begin{eqnarray}}
\def\ear{\end{eqnarray}}

\def\eqi{\begin{equation}}
\def\eqf{\end{equation}}
\def\eqia{\begin{eqnarray}}
\def\eqfa{\end{eqnarray}}
\def\rp#1#2{{#1\over#2}}
\def\ct#1{\cite{#1}}
\def\lb#1{\label{#1}}

\def\oc2{$\mathcal{O}(c^{-2})$}

\documentclass[11pt]{article}
\usepackage{amsmath,amsthm,amscd,amssymb}
\usepackage{latexsym}
\usepackage{graphicx,epsfig}

\begin{document}

\noindent{\bf \LARGE{Solar System planetary orbital motions and
dark matter}}
\\
\\
\\
{L. Iorio, \textit{FRAS}}\\
{\it Viale Unit$\grave{a}$ di Italia 68, 70125\\Bari, Italy
\\tel./fax 0039 080 5443144
\\e-mail: lorenzo.iorio@libero.it}

\begin{abstract}
In this paper we explicitly work out the effects that a
spherically symmetric distribution of dark matter with constant
density  would induce on the Keplerian orbital elements of the
Solar System planets and compare them with the latest results in
planetary orbit determination from the EPM2004 ephemerides. It
turns out that the longitudes of perihelia $\varpi$ and the mean
longitudes $\lambda$ are affected by secular precessions. The
resulting upper bounds on dark matter density, obtained from the
EPM2004 formal errors in the determined mean longitude shifts over
90 years, lie in the range $10^{-19}-10^{-20}$ g cm$^{-3}$ with a
peak of $10^{-22}$ g cm$^{-3}$ for Mars. Suitable combinations of
the planetary mean longitudes and perihelia, which cancel out the
aliasing impact of some of the unmodelled or mismodelled forces of
the dynamical models of EPM2004, yield a global upper bound of
$7\times 10^{-20}$ g cm$^{-3}$ and $4\times 10^{-19}$ g cm$^{-3}$,
respectively.
\end{abstract}

\\
\section{Introduction}
In this paper we exploit the latest accurate data on planetary
orbit determination \ct{pitSSR} in order to constraint the amount
of dark matter present in the Solar System. To avoid possible
misunderstandings, here we are not intending the amount of dark
matter at the Solar System's position in the Galaxy; we are,
instead, dealing with possible excess of dark matter above the
galactic background density\footnote{Such a galactic background
density would not cause the perturbing effects on planetary
motions discussed in this paper; it is believed to cause an
overall acceleration toward the center of the Galaxy.}. Such
excess is to be intended as coming from dark matter
gravitationally bound to the Solar System and partly from dark
matter in the galactic halo focused by the solar gravitational
potential.

Let us consider a distribution with spherically symmetric density
$\rdm(r)$.  The corresponding modification to the Newtonian
gravitational potential can be cast in the form \eqi U_{\rm
dm}=-\rp{G\mu(r)}{r},\lb{udm}\eqf where
\eqi\mu(r)=4\pi\int_{0}^{r}\rdm(r^{'})r^{'2}dr^{'}\eqf is the
total mass of dark matter enclosed in a spherical region of radius
$r$. In regard to the distribution of the dark matter, it is
believed it comes from diffusion into the Solar System (or
possibly it was already present when our planetary system was
formed). Diffusion was studied, e.g. in \ct{pippo, pluto}, but it
turns out that in none of these papers the density distribution in
the Solar System has been derived in a form suitable for the
present paper. Thus, we will make the simplest choice, i.e. we
will consider $\rdm$ as constant. With this assumption
$\mu=(4/3)\pi\rdm r^3$ and \rfr{udm} becomes \eqi U_{\rm dm
}=-\rp{4}{3}\pi G\rdm r^2.\eqf From it an entirely radial
acceleration \eqi {\boldsymbol{A}}_{\rm dm}=\rp{8}{3}\pi G\rdm
\boldsymbol{r}\lb{ADM}\eqf occurs. Its effect on planetary
motions, which is, of course, much smaller than usual Newtonian
gravity, can straightforwardly be calculated within the usual
perturbative schemes.
\section{The impact of a spherically symmetric distribution of dark matter on the planetary orbits}
The Gauss equations for the variations of the semimajor axis $a$,
the eccentricity $e$, the inclination $i$, the longitude of the
ascending node $\Omega$, the argument of pericentre $\omega$ and
the mean anomaly $\mathcal{M}$ of a test particle in the
gravitational field of a body $M$ are
\bar
\dert{a}{t} & = & \rp{2}{n\sqrt{1-e^2}}\left[eA_r\sin f+A_t\left(\rp{p}{r}\right)\right],\lb{smax}\\
\dert{e}{t} & = & \rp{\sqrt{1-e^2}}{na}\left\{A_r\sin f+A_t\left[\cos f+\rp{1}{e}\left(1-\rp{r}{a}\right)\right]\right\},\\
\dert{i}{t} & = & \rp{1}{na\sqrt{1-e^2}}\ A_n\left(\rp{r}{a}\right)\cos (\omega+f),\lb{in}\\
\dert{\Omega}{t} & = & \rp{1}{na\sin i\sqrt{1-e^2}}\ A_n\left(\rp{r}{a}\right)\sin (\omega+f),\lb{nod}\\
\dert{\omega}{t} & = & -\cos i\dert{\Omega}{t}+\rp{\sqrt{1-e^2}}{nae}\left[-A_r\cos f+A_t\left(1+\rp{r}{p}\right)\sin f\right],\lb{perigeo}\\
\dert{\mathcal{M}}{t} & = & n -\rp{2}{na}\
A_r\left(\rp{r}{a}\right)-\sqrt{1-e^2}\left(\dert{\omega}{t}+\cos
i\dert{\Omega}{t}\right),\lb{manom} \ear
in which $n=2\pi/P$ is the  mean motion\footnote{For an
unperturbed Keplerian ellipse it amounts to $n=\sqrt{GM/a^3}$.},
$P$ is the test particle's orbital period, $f$ is the true anomaly
counted from the pericentre, $p=a(1-e^2)$ is the semilactus rectum
of the Keplerian ellipse, $A_r,\ A_t,\ A_n$ are the  radial,
transverse (in-plane components) and the normal (out-of-plane
component) projections of the perturbing acceleration
$\boldsymbol{A}$, respectively, on the frame
$\{\boldsymbol{\hat{r}},\boldsymbol{\hat{t}},\boldsymbol{\hat{n}}\}$
comoving with the particle.

If the perturbing acceleration is entirely radial, as it is the
case for \rfr{ADM}, the Gauss equations reduce to

\bar
\dert{a}{t} & = & \rp{2e}{n\sqrt{1-e^2}}\ A_r\sin f,\lb{smax2}\\
\dert{e}{t} & = & \rp{\sqrt{1-e^2}}{na}\ A_r\sin f,\\
\dert{i}{t} & = & 0,\lb{in2}\\
\dert{\Omega}{t} & = & 0,\lb{nod2}\\
\dert{\omega}{t} & = & -\rp{\sqrt{1-e^2}}{nae}\ A_r\cos f,\lb{perigeo2}\\
\dert{\mathcal{M}}{t} & = & n -\rp{2}{na}\
A_r\left(\rp{r}{a}\right)-\sqrt{1-e^2}\
\dert{\omega}{t},\lb{manom2} \ear

As a result, the inclination and the node are not perturbed by
such a perturbing acceleration.

By evaluating \rfr{ADM} on the unperturbed Keplerian ellipse \eqi
r=\rp{a(1-e^2)}{1+e\cos f},\eqf inserting it into
\rfrs{smax2}{manom2} and averaging them with \eqi
\rp{dt}{P}=\rp{1}{2\pi}\rp{(1-e^2)^{3/2}}{(1+e\cos f)^2}df \eqf
one gets that only the argument of pericentre and the mean anomaly
are affected by extra secular precessions \bar
\left\langle\dert\omega t
\right\rangle &=&\rp{4\pi G\rdm}{n}\sqrt{1-e^2},\label{dodt}\\
 \left\langle\dert{\mathcal{M}} t
\right\rangle &=&-\rp{4\pi G\rdm
}{n}\left(\rp{7}{3}+e^2\right).\ear As a consequence, also the
mean longitude $\lambda=\Omega+\omega+\mathcal{M}$, which is used
for orbits with small inclinations and eccentricities as those of
the Solar System planets, is affected by an extra secular rate
\eqi\left\langle\dert\lambda t \right\rangle=\rp{4\pi
G\rdm}{n}\left(\sqrt{1-e^2}-\rp{7}{3}-e^2\right).\label{dldt}\eqf
\section{Confrontation with the latest data}
\subsection{The longitudes of perihelia}\lb{perihl}
The obtained expressions are useful for comparison with the latest
data on planetary orbits from the EPM2004 ephemerides \ct{pitSSR}.
A similar analysis was conducted in \ct{KhriPit} by exploiting the
determined extra-advances $\Delta\dot\varpi$ of the longitudes of
perihelia $\varpi$ of the inner planets \ct{pitAL}. By using the
uncertainties  $\delta(\Delta\dot\varpi)$ released in \ct{pitAL},
the authors of \ct{KhriPit} obtained an upper bound on $\rdm$ of
$10^{-17}$ g cm$^{-3}$ from Mercury and of $10^{-19}$ g cm$^{-3}$
from the Earth and Mars. Note that \rfr{dodt} yields for the
perihelion advance, in units of $2\pi$, after one full orbital
revolution  the same expression (5) of \ct{KhriPit}, apart from
the eccentricity factor. Moreover, the constraints on $\rdm$
obtainable for each planet separately  with \rfr{dodt} and the
results of \ct{pitAL} fully agree with those of \ct{KhriPit}.
However, the following  concern about the reliability of the
estimates by the authors of \ct{KhriPit} is in order. The
extra-perihelion advances determined in \ct{pitAL} are affected,
in general, by all the Newtonian and non-Newtonian features of
motion which have not been accounted for in the dynamical force
models of EPM2004. Among them there are certainly the totally
umodelled general relativistic gravitomagnetic field, which
induces the Lense-Thirring planetary precessions, and the solar
quadrupole mass moment $J_2$ which, instead, was included in
EPM2004, but it is currently affected by a $\sim 10\%$
uncertainty. For a discussion of such issues in a different
context see \ct{iorlt}. Each planetary perihelion is, thus,
affected by such residual effects, so that it is not possible to
entirely attribute the determined extra-advances to the action of
the dark matter distribution, especially for Mercury. A better way
to use $\Delta\dot\varpi$ is to suitably combine them in order to
make the estimate of $\rdm$ independent, by construction, of the
Lense-Thirring and $J_2$ effects. According to the approach
followed in \ct{iorlt, ioraa}, it is possible to construct the
following combination \eqi\rdm=\rp{\Delta\dot\varpi^{\rm
Mer}+c_1\Delta\dot\varpi^{\rm Ear}+c_2\Delta\dot\varpi^{\rm
Mar}}{1.7\times 10^{10}\ {\rm cm}^3\ {\rm yr}^{-1}\ {\rm g}^{-1}}
,\lb{combi1}\eqf with $c_1=-80.7$ and $c_2=217.6$. The
dimensionless coefficients $c_1$ and $c_2$, which are built up
with $a,e$ and $i$ of the planets adopted, cancel out, by
construction, the impact of the gravitomagnetic field and of the
solar quadrupolar mass moment on the combination \rfr{combi1}.
This can straightforwardly be checked by combining the perihelion
precessions due to $J_2$ and to the Lense-Thirring force with the
coefficients of \rfr{combi1}: the result is zero. The upper bound
on $\rdm$ obtainable from \rfr{combi1} can conservatively be
obtained from\footnote{Although rather small, a certain amount of
correlation among the determined extra-advances of perihelia is
present, with a maximum of $20\%$ between the Earth and Mars (E.V.
Pitjeva, private communication, 2006). If a root-sum-square
calculation is performed, an upper bound $\rdm\leq 3\times
10^{-19}$ g cm$^{-3}$ is obtained.} \eqi\delta\rdm\leq
\rp{\delta(\Delta\dot\varpi^{\rm
Mer})+|c_1|\delta(\Delta\dot\varpi^{\rm
Ear})+|c_2|\delta(\Delta\dot\varpi^{\rm Mar})}{1.7\times 10^{10}\
{\rm cm}^3\ {\rm yr}^{-1}\ {\rm g}^{-1}};\eqf it amounts to
$4\times 10^{-19}$ g cm$^{-3}$.
\subsection{The mean longitudes}

Here we compare the shifts in $\lambda$ obtainable from \rfr{dldt}
with the formal errors in fitting the mean longitudes at epoch of
all the nine major planets by using a data record spanning 90
years from Table 4 of \ct{pitSSR}. Our results are in Table
\ref{tavola}.
{\small\begin{table}\caption{Upper bounds, in g cm$^{-3}$, on
$\rdm$ from the comparison of the planetary mean longitude shifts
over 90 years from \rfr{dldt} and the formal errors in the
planetary mean longitudes released in \ct{pitSSR}. }\label{tavola}
\begin{center}
\begin{tabular}{cc}
\hline
 Mercury & $4\times 10^{-19}$\\
 Venus  & $9\times 10^{-20}$\\
 Mars & $5\times 10^{-22}$\\
 Jupiter & $2\times 10^{-20}$\\
 Saturn & $3\times 10^{-20}$\\
 Uranus & $3\times 10^{-20}$\\
  Neptune & $6\times 10^{-20}$\\
  Pluto & $9\times 10^{-20}$\\
\hline
%
%
\end{tabular}
\end{center}
\end{table}}
As can be noted, the tightest constraint comes from Mars, with
$5\times 10^{-22}$ g cm$^{-3}$. The present upper bounds are
better than those obtained in \ct{KhriPit} from the perihelia of
each planet, even if we conservatively assume that the realistic
errors in $\lambda$ may be one order of magnitude larger than
those released in \ct{pitSSR}. In regard to the constraint from
Uranus $3\times 10^{-20}$ g cm$^{-3}$, the previously obtained
upper bound on $\rdm$ from such a planet amounted to $10^{-16}$ g
cm$^{-3}$ \ct{AndJPL}. From Table \ref{tavola2} it can be noted
that with a such high value of $\rdm$ the periehelia and the mean
longitudes of the outer planets of Solar System would undergo very
huge secular precessions up to tens of arcseconds per century.
Such anomalous effects would have certainly been detected, even in
the case of the outer planets whose orbital determination relies
almost entirely upon relatively inaccurate  optical data
\ct{pitSSR}.
%
%
%
{\small\begin{table}\caption{Secular precessions of longitudes of
perihelia and mean longitudes, in arcseconds per century, induced
on the outer planets by a dark matter density $\rdm=10^{-16}$ g
cm$^{-3}$ equal to the upper bound found in \ct{AndJPL}.
}\label{tavola2}
\begin{center}
\begin{tabular}{ccc}
\hline
Planet & $\dot\varpi$ & $\dot\lambda$ \\
\hline
Jupiter & 3.25 & -4.349\\
 Saturn & 8.06 & -10.803\\
 Uranus & 23.0 & -30.807\\
  Neptune & 45 & -60.27\\
  Pluto & 65.8 & -97.03\\
\hline
%
%
\end{tabular}
\end{center}
\end{table}}

However, caution is advised in considering the results of Table
\ref{tavola} for each planet separately. It is so not only for the
same reasons already exposed in Section \ref{perihl} about the
unmodelled or mismodelled forces, but also because of the
systematic errors $\delta n=(3/2)\sqrt{GM/a^5}\delta a$ in the
Keplerian mean motions induced by the uncertainties in the
semimajor axes\footnote{ I am grateful to M. Sereno who pointed
out this fact to me.}. Following the approach of Section
\ref{perihl}, we can write down the combination
\eqi\rdm=\rp{\Delta\dot\lambda^{\rm Mer}+k_1\Delta\dot\lambda^{\rm
Ven}+k_2\Delta\dot\lambda^{\rm Mar}}{-2.4\times 10^{10}\ {\rm
cm}^3\ {\rm yr}^{-1}\ {\rm g}^{-1}} ,\lb{combi2}\eqf with
$k_1=-15.4$ and $k_2=78.1$. Also such a combination is independent
of the solar Lense-Thirring and $J_2$ effects. The formal errors
in $\lambda$ of Table 4 of \ct{pitSSR}, combined according to
\rfr{combi2}, yields a conservative upper bound $\delta\rdm\leq
7\times 10^{-21}$ g cm$^{-3}$. The errors in $n$, linearly
combined according to \rfr{combi2}, yield a systematic bias
$\delta\rdm\leq 7.7\times 10^{-20}$ g cm$^{-3}$, so that a total
upper bound $\delta\rdm\leq 7.8\times 10^{-20}$ g cm$^{-3}$ can be
obtained with the squared sum of these two kinds of errors. As can
be noted, the use of \rfr{combi2} yields better results that
\rfr{combi1}.
\section{Conclusions}
In this paper we have worked out the effects that a local excess
of dark matter in our Solar System over the galactic background
would induce on the orbits of the planets. It turns out that, by
modelling the dark matter distribution with a constant density,
the longitude of perihelion and the mean longitude of a planet are
affected by secular precessions. The comparison with the latest
data show that the upper bounds obtainable from the mean
longitudes and the perihelia are of the order of $10^{-20}$ g
cm$^{-3}$ and $10^{-19}$ g cm$^{-3}$, respectively.

\section*{Acknowledgements}
I thank I. Khriplovich, E.V. Pitjeva and M. Sereno for stimulating
correspondence and useful observations. I am also grateful to the
anonymous referee for his important remarks who greatly
contributed to the improvement of the present paper.


\end{document}